\documentclass[useAMS,usenatbib]{mn2e}
\usepackage{graphicx}
\usepackage{subfigure}
\usepackage{rotating} 
\usepackage{float}

\newcommand{\lsim }{{\lower0.8ex\hbox{$\buildrel <\over\sim$}}}
\newcommand{\gsim }{{\lower0.8ex\hbox{$\buildrel >\over\sim$}}}

\usepackage{soul}

\def\chandraacis{\emph{Chandra/ACIS}}

\def\simge{\mathrel{%
  \rlap{\raise 0.511ex \hbox{$>$}}{\lower 0.511ex \hbox{$\sim$}}}}
\def\simle{\mathrel{
  \rlap{\raise 0.511ex \hbox{$<$}}{\lower 0.511ex \hbox{$\sim$}}}}

\newcommand{\Msun}{\ifmmode {M_{\odot}}\else${M_{\odot}}$\fi}
\newcommand{\Lsun}{\ifmmode {L_{\odot}}\else${L_{\odot}}$\fi}
\newcommand{\Rsun}{\ifmmode {R_{\odot}}\else${R_{\odot}}$\fi}

\usepackage{xargs}
\usepackage{color}

\title[Thermal $L_X$ variations in qLMXBs]{Limits on thermal variations in a dozen quiescent neutron stars over a decade}
\author[Bahramian et al.]{
Arash Bahramian$^{1}$\thanks{E-mail: bahramia@ualberta.ca},
Craig O. Heinke$^{1,2}$,
Nathalie Degenaar$^{3}$,
Laura Chomiuk$^{4}$,
\newauthor Rudy Wijnands$^{5}$,
Jay Strader$^{4}$,
Wynn C. G. Ho$^6$,
David Pooley$^{7,8}$
\\
$^{1}$ Dept. of Physics, CCIS 4-183, University of Alberta, Edmonton, AB T6G 2E1, Canada\\
$^{2}$ Humboldt Fellow, Max Planck Institut f$\ddot{u}$r Radioastronomie, Auf dem H$\ddot{u}$gel 69, 53121 Bonn, Germany\\
$^{3}$ Institute of Astronomy, University of Cambridge, Madingley Road, Cambridge CB3 OHA, UK\\
$^{4}$ Department of Physics and Astronomy, Michigan State University, East Lansing, MI 48824, USA\\
$^{5}$ Anton Pannekoek Institute for Astronomy, University of Amsterdam, Postbus 94249, NL-1090 GE Amsterdam, The Netherlands\\
$^{6}$ Mathematical Sciences and STAG Research Centre, University of Southampton, Southampton, SO17 1BJ, United Kingdom\\
$^{7}$ Eureka Scientific, 2452 Delmer Street, Suite 100, Oakland, CA 94602-3017, USA\\
$^{8}$ Sam Houston State University, Department of Physics, Farrington Building, Suite 204, Huntsville, Texas 77341, USA}
\begin{document}

\date{}

\pagerange{\pageref{firstpage}--\pageref{lastpage}} \pubyear{2015}

\maketitle

\label{firstpage}

\begin{abstract}
In quiescent low-mass X-ray binaries (qLMXBs) containing neutron stars, the origin of the thermal X-ray component may be either release of heat from the core of the neutron star, or continuing low-level accretion. In general, heat from the core should be stable on timescales $<10^4$ years, while continuing accretion may produce variations on a range of timescales. While some quiescent neutron stars (e.g. Cen X-4, Aql X-1) have shown variations in their thermal components on a range of timescales, several others, particularly those in globular clusters with no detectable nonthermal hard X-rays (fit with a powerlaw), have shown no measurable variations. 
Here, we constrain the spectral variations of 12 low mass X-ray binaries in 3 globular clusters over $\sim10$ years. We find no evidence of variations in 10 cases, with limits on temperature variations below 11\% for the 7 qLMXBs without powerlaw components, and limits on variations below 20\% for 3 other qLMXBs that do show non-thermal emission. However, in 2 qLMXBs showing powerlaw components in their spectra (NGC 6440 CX 1 \& Terzan 5 CX 12) we find marginal evidence for a 10\% decline in temperature, suggesting the presence of continuing low-level accretion. This work adds to the evidence that the thermal X-ray component in quiescent neutron stars without powerlaw components can be explained by heat deposited in the core during outbursts. Finally, we also investigate the correlation between hydrogen column density (N$_H$) and optical extinction (A$_V$) using our sample and current models of interstellar X-ray absorption, finding $N_H ({\rm cm}^{-2}) = (2.81\pm0.13)\times10^{21} A_V$.
\end{abstract}


\section{Introduction}
Transient low-mass X-ray binaries (LMXBs) show occasional outbursts, separated by periods of quiescence (typically months to years) in which their X-ray emission dramatically drops, to $L_X=10^{31}$--$10^{33}$ erg s$^{-1}$, and there is little or no accretion occurring. The X-ray spectra of quiescent LMXBs (qLMXBs) containing a neutron star (NS) consist of a soft blackbody-like component (if fit with a blackbody, giving typical temperatures of 0.2-0.3 keV) and sometimes an additional harder component, typically fit with a powerlaw, with photon-index between 1 and 2 \citep[e.g.][]{Campana98a,Rutledge99}. The origin of the powerlaw component in these systems may be continuing low level accretion onto the NS, an accretion shock between infalling matter and the NS's magnetosphere or a pulsar wind, or synchrotron radiation from a pulsar wind nebula \citep[e.g.][]{Campana98a,Bogdanov05}.  Recent X-ray observations support boundary-layer emission from accretion onto the NS as the likely origin of the hard powerlaw component at least in some sources (e.g. Cen X-4 and PSR J1023+0038, \citealt{Chakrabarty14,D'Angelo14,Archibald15,Wijnands14}). 

The blackbody-like component can be well-described by NS hydrogen atmosphere models, with an implied radius consistent with theoretically predicted NS radii \citep[e.g.][]{Rajagopal96,Zavlin96,Rutledge99,Rutledge01a}. The soft component has been widely interpreted as the slow ($10^4$ years) release of heat from the core of the NS, deposited through deep crustal heating during previous episodes of accretion \citep{Brown98}. A nearly identical soft spectrum can also be created in the upper layers of the NS atmosphere by low-level accretion \citep{Zampieri95,Zane00}, though the most detailed model predicts, in addition, an optically thin high-temperature bremsstrahlung component \citep{Deufel01}, which may be identified with the powerlaw component \citep[e.g.][]{Bahramian14}.  However, low-level accretion at sufficiently high rates (corresponding to $L\sim10^{33}$ erg/s) could maintain metals in the atmosphere at abundances sufficient to soften the spectrum, leading to an overestimate of R \citep{Rutledge02b}.
The question of the origin of the thermal component in qLMXBs is of intense interest, since fitting the thermal spectra of qLMXBs in globular clusters is often used to measure the radius of NSs and place constraints on the dense matter equation of state (\citealt{Rutledge02a,Heinke06a,Webb07,Steiner10,Steiner13,Guillot11,Servillat12,Guillot13,Guillot14,Lattimer14,Heinke14,Ozel15}; see also \citealt{Miller13}). Confirmation that thermal spectra of qLMXBs (especially those without powerlaw components) are not produced by accretion would eliminate the possibility that metals remain in the atmosphere, and thus eliminate a systematic uncertainty in this method of constraining the NS radius.

At low accretion rates, the NS's magnetic field is expected to exercise a propeller effect, which retards the accretion of material onto the NS surface \citep{Illarionov75}. The mechanism is that ionized disc material will become attached to the magnetic field lines when the magnetospheric pressure exceeds the gas pressure; if this boundary (the edge of the magnetosphere) occurs at a radius larger than the corotation radius, the material is accelerated to velocities higher than the Keplerian orbital velocity, throwing it away, and possibly out of the system.
On the other hand, simulations of low-level accretion onto a magnetic propeller predict some material will reach the neutron star \citep{Romanova02,Kulkarni08,D'Angelo10,D'Angelo12}. 
Pulsations have recently been identified from two (relatively bright) qLMXBs, the transitional pulsars PSR J1023+0038 at $L_X$=$3\times10^{33}$ erg/s, and XSS J12270-4859 at $L_X$=$5\times10^{33}$ erg/s, proving that matter accretes onto the surface even though it seems to be in the propeller regime \citep{Archibald15,Pappito15}.
Detections of pulsations at low accretion rates from X-ray pulsars (with known magnetic fields) in high-mass X-ray binaries also suggest continued accretion in the propeller regime  \citep[e.g.][]{Negueruela00,Reig14}. Note that pulsations have not yet been detected from quiescent LMXBs other than transitional MSPs (\citealt{D'Angelo14}; Elshamouty et al., in prep.).

Thus, there is evidence that continued accretion can produce an X-ray spectrum consistent with the soft component in qLMXBs, by heating the NS surface \citep{Zampieri95,Zane00,Deufel01} and that continued accretion can continue at very low accretion rates. Testing for variability is a promising method to constrain whether thermal emission from qLMXBs is driven by accretion.  
X-ray emission driven by heat emerging from the core should be stable on timescales $\sim10^4$ years \citep{Colpi01,Wijnands13}.  For NSs that have undergone recent episodes of accretion, decays of the thermal component are seen \citep[e.g.][]{Wijnands02b,Cackett08,Degenaar11exo0748,Fridriksson11}, attributed to heat leaking from shallower levels in the crust, where it was deposited during the outburst \citep{Ushomirsky01,Rutledge02c}.  This variation should be a monotonic decline on a time scale of years/decades. Possibilities for variations of the thermal component without accretion, based on changes in the chemical composition of the envelope, have been advanced \citep{Brown02,Chang04}, but these do not predict significant changes on timescales of years without intervening outbursts.
Alternatively, if the thermal component is caused by continuous accretion onto a NS, the temperature may vary in either direction, on a range of timescales. 

Strong evidence for variability of the thermal component of some qLMXBs during quiescence has been reported for the transients Aquila X-1 \citep{Rutledge02b,Cackett11}, Cen X-4 \citep{Campana97,Campana04a,Cackett10}, XTE J1701-462 \citep{Fridriksson10}, and MAXI J0556-332 \citep{Homan14}. 
Each of the transient qLMXBs that showed strong variation in the thermal component also showed a relatively strong powerlaw component, typically making up $\sim$50\% of the 0.5-10 keV flux \citep[e.g.][]{Rutledge02b,Cackett10,Homan14}. 
So far, this is consistent with the suggestion \citep{Heinke03d} that continued accretion is responsible for both the powerlaw component (at least for $L_X\gsim10^{33}$ erg/s, \citealt{Jonker04b}) and variability.

Globular clusters (GCs) are highly efficient factories for producing X-ray sources \citep[$\sim$100 times more efficient than the rest of the Galaxy, per unit mass;][]{Katz75,Clark75}.The densest and most massive globular clusters have been shown by Chandra to contain multiple soft X-ray sources with $L_X$ in the $10^{32}$--$10^{33}$ erg/s range, whose spectra strongly indicate that they are qLMXBs \citep{Grindlay01a,Pooley02b,Heinke03b,Heinke03c,Pooley03,Maxwell12}.  These can be discriminated from other sources (spectrally harder cataclysmic variables, chromospherically active binaries, and millisecond radio pulsars) through spectral fitting \citep[e.g.][]{Rutledge02a}, or through X-ray colours and luminosities \citep{Heinke03d,Pooley06}.
The majority of the candidate qLMXBs show no evidence for powerlaw components in their spectra \citep{Heinke03d}.  This may be a selection effect, since it is easier to discriminate qLMXBs from other sources if they show purely thermal spectra. Deep Chandra observations of globular clusters identify a significant population of candidate qLMXBs with strong powerlaw components, which require high-quality X-ray spectra to confidently identify the thermal components \citep{Heinke05b}, and half of the transient cluster LMXBs followed up with Chandra are too faint, or too dominated by a powerlaw component, for detection of a thermal component \citep[e.g.][]{Wijnands05,Heinke10,Linares14}.  

Several qLMXBs in globular clusters with thermal components have been examined to search for variations between observations.  Several show flux variations, but spectral analyses have permitted these variations to be driven only by changes in the normalization of the powerlaw component \citep{Heinke05b,Cackett05,Bahramian14}, or of obscuring material--the latter especially in edge-on systems \citep[e.g.][]{Nowak02,Heinke03a,Wijnands03a}.  
A few qLMXBs without detectable powerlaw components have repeated deep observations with {\it Chandra}, permitting sensitive measurements for variability; these include (with 90\% confidence variability limits) 47 Tuc X7, $\Delta T/T<$1.0\% over 2 years \citep{Heinke06a}; NGC 6397 U18, $\Delta T/T<$1.4\% over 10 years \citep{Guillot11,Heinke14}; M28 source 26 \citep[no variation found,][]{Servillat12} 
and $\omega$ Cen, $\Delta T/T<$2.1\% over 12 years \citep{Heinke14}.  

We are interested in whether this lack of variability is the norm in thermal components of globular cluster qLMXBs, and whether (if it is observed anywhere) there is a correlation with the presence of powerlaw components. Strong evidence that globular cluster qLMXBs without powerlaw components are not variable would increase confidence in the assumptions used to derive radius measurements for their NSs and obtain constraints on the dense matter equation of state. We choose to focus on qLMXBs in globular clusters, since globular clusters are the only places where we can find and study large populations of qLMXBs in single Chandra pointings (and with relatively low optical extinction, compared to many qLMXBs in the Galactic Plane) and distances to qLMXBs in globular clusters are known better than to the ones in the field.

In this paper, we analyze a sample of 12 qLMXBs in 3 globular clusters, NGC 6266 (M 62), NGC 6440, and Terzan 5, searching for variations over multiple observations spanning roughly a decade. Each of these dense globular clusters have at least 4 candidate qLMXBs, and multiple \chandraacis\ observations. When this work was in an advanced state, \citet{Walsh15} reported on the analysis of 9 similar sources in Terzan 5 and NGC 6440 as considered here, though using fewer Terzan 5 observations, and not including NGC 6266. We directly compare our results with those of Walsh et al., finding general agreement. In \S \ref{sec_data}, we describe our data reduction and analysis methods. In \S \ref{sec_result}, we show the results of our analysis, and in \S \ref{sec_disc}, we discuss the implications.

\section{Data reduction and Analysis}\label{sec_data}
All datasets were obtained using ACIS-S in Faint, Timed Exposure mode. The data covers a time span of 9 to 12 years for the GCs in our sample (Table~\ref{tab_data}). We used CIAO 4.6 \citep{Fruscione06} and CALDB 4.6.2 for data reprocessing following standard CIAO science threads\footnote{http://cxc.harvard.edu/ciao/threads/index.html}. We chose known (and candidate) LMXBs with more than $\sim$ 60 photons (per epoch) for spectral analysis, resulting in a total sample of 12 sources. 
These targets are tabulated in Table \ref{tab_targets}. We performed spectral extraction using the task \emph{specextract} and performed spectral analysis in the 0.3-10 keV energy range using HEASOFT 6.16 and XSPEC 12.8.2 \citep{Arnaud96}. We combined (using the CIAO \emph{dmmerge} tool) spectra from observations which occurred within a month for our targets in Terzan 5, to improve the spectral quality. We grouped each spectrum by 15 counts per bin, and used chi-squared statistics for analysis. Details of the spectral extractions and analyses for each cluster are discussed in \S \ref{sec_result}. All uncertainties reported in this paper are 90\% confidence.

We fit the spectra with an absorbed neutron star atmosphere, plus a powerlaw (if needed).
 We used the NSATMOS \citep{Heinke06a} NS atmosphere model, with the NS mass and radius frozen to the canonical values of 1.4 M$_\odot$ and 10 km, the normalization fixed to 1 (implying radiation from the entire surface), and the distance fixed to the estimated distance of the cluster (see \S \ref{sec_result}). In cases where this single-component model does not produce a good fit, leaving significant residuals at high energies($\geq$3 keV), a powerlaw component (PEGPWRLW, with normalization set to represent flux in the 0.5-10 keV range) was added to the model. In cases where it was unclear whether the fit was improved by adding a second component, we checked this by an F-test. We set the powerlaw photon index ($\Gamma$) as a free parameter when fitting datasets where the data quality above 3 keV is high enough to constrain this quantity. However, in datasets where the data quality is not sufficient to do this, we fixed the photon index to 1.5, which is a typical value from LMXBs in quiescence \citep{Campana98a,Campana04a,Fridriksson11,Degenaar11exo0748,Chakrabarty14}.

We fit all spectra available for each source simultaneously. We tied the N$_H$ and powerlaw photon index to a single value for each source, and let the NS temperature and powerlaw flux vary between observations.  There is good evidence that N$_H$ does not vary during the spectral evolution of most LMXBs \citep{Miller09}. Although correlated variations in N$_H$ and powerlaw photon index have been proposed to explain spectral variability in Aql X-1 in quiescence \citep{Campana03}, such a model requires the powerlaw to dominate over the thermal component of the spectrum, and forces the photon index to (unreasonably) high values ($\geq2.5$).  Fitting individual high-quality spectra of Aql X-1 \citep{Cackett11}, SAX J1808.4-3658 \citep{Heinke09a}, and Terzan 5 X-3 \citep{Bahramian14} in quiescence gives consistent values (within the errors) between epochs for both N$_H$ and the powerlaw photon index. Deep observations of Cen X-4 do show evidence for powerlaw photon index changes \citep{Cackett10}, changing from $\sim$1.7 in all observations where $L_X<2\times10^{32}$ erg/s, to $\sim$1.4 at higher $L_X$ values \citep{Chakrabarty14}.  \citet{Fridriksson10} also find evidence for variation in the powerlaw photon index in XTE J1701-462, but only when $L_X$ reaches $>2\times10^{34}$ erg/s, well above the range discussed here.  Summarizing these observations, when the total $L_X$ does not vary by a large factor, the powerlaw photon index appears to remain roughly constant.  As we will see below, our observations do not reveal large swings in $L_X$, so we feel that keeping a fixed powerlaw index is justified.

As we are looking for relative variations in $kT$, to estimate the uncertainties in the NS temperature and powerlaw flux, after initially fitting the $N_H$ we then froze N$_H$ to its best-fit value. In cases where the source shows significantly higher N$_H$ value than the average of the cluster, we investigated the possibility of intrinsic absorption leading to variations in N$_H$ (\S \ref{sec_result}).

We also search for signs of temporal variations during each observation for all sources.  (Due to the limited signal, we explore only total flux variations.) We extracted background-subtracted lightcurves in 0.3-7 keV band for each source using CIAO task \emph{dmextract}. Due to the faint nature of the sources in our study, we binned each lightcurve by 1000 s. We used the FTools task \emph{lcstats} to perform chi-squared and Kolmogorov-Smirnov (KS) tests for variations. Since we ran these tests on 80 lightcurves (all sources in our sample in all observations used\footnote{We excluded one of Terzan 5 observations (Obs.ID 13705) from this investigation due to its short exposure (14 ks).}), we require a false-alarm probability below 5\%/80=0.06\% to identify variability at 95\% confidence (Note that null-results from a test do not prove the source did not vary).

\begin{table*}
\centering
\begin{tabular}{@{}cllllllc@{}}
\hline
Target		&	Obs. ID	&	Date			&	Exposure (ks)\\
\hline
NGC 6266	&	02677	&	2002-05-12	&	62\\
			&	15761	&	2014-05-05	&	82\\
\\
NGC 6440	&	00947	&	2000-07-04 	&	23\\
			&	03799	&	2003-06-27	&	24\\
			&	10060	&	2009-07-28	&	49\\
\\
Terzan 5		&	03798	&	2003-07-13	&	40\\	
			&	10059	&	2009-07-15 	&	36\\		
			&	13225	&	2011-02-17	&	30\\		
			&	13252	&	2011-04-29	&	40\\		
			&	13705	&	2011-09-05	&	14$^A$ \\		
			&	14339	&	2011-09-08 	&	34$^A$ \\		
			&	13706	&	2012-05-13	&	46\\		
			&	14475	&	2012-09-17	&	30\\		
			&	14476	&	2012-10-28 	&	29\\		
			&	14477	&	2013-02-05	&	29$^B$\\		
			&	14625	&	2013-02-22	&	49$^B$\\		
			&	15615	&	2013-02-23 	&	84$^B$\\		
			&	14478	&	2013-07-16	&	29\\		
			&	14479	&	2014-07-15	&	29$^C$\\
			&	16638	&	2014-07-17	&	72$^C$\\	
			&	15750	&	2014-07-20	&	23$^C$\\		
\hline
\end{tabular}
\caption{Chandra ACIS observations used in this study. A, B, C: We merged spectra extracted from all observations marked A (and merged those marked B, etc.), as they occurred close in time.}
\label{tab_data}
\end{table*}

\begin{table*}
\centering
\begin{tabular}{@{}lccl@{}}
\hline
Source			&	CXOGlb J			&L$_X$ 	\\
				&	&(10$^{32}$erg s$^{-1}$, 0.5-10 keV) \\
\hline
NGC 6266 CX4		& 170113.09--300655.43	&	4.5$\pm0.3$	\\
NGC 6266 CX5		& 170113.10--300642.33	&	4.2$\pm0.3$	\\
NGC 6266 CX6		& 170113.76--300632.48	&	3.9$\pm0.3$	\\
NGC 6266 CX16	& 170112.58--300622.38	&	1.0$\pm0.2$	\\
\hline
NGC 6440 CX1		& 		--			&	10$\pm1$		\\
NGC 6440 CX2		& 		--			&	14$\pm1$		\\
NGC 6440 CX3		& 		--			&	8$\pm1$		\\
NGC 6440 CX5		& 		--			&	7.4$\pm0.8$	\\
\hline
Terzan 5 CX9		&  174804.8--244644	&	9.0$\pm0.9$	\\
Terzan 5 CX12		&  174806.2--244642	&	7.0$\pm0.8$	\\
Terzan 5 CX18		&  174805.2--244651	&	6.3$\pm0.6$	\\
Terzan 5 CX21		&  174804.2--244625	&	4.4$\pm0.5$	\\
\hline
\end{tabular}
\label{tab_targets}
\caption{Candidate LMXBs studied in this paper. Luminosities are based on latest epoch used for each GC and are measured in this study. Reported uncertainties for luminosities are 90\% confidence and do not include uncertainties in distance. References for source identification: This work (NGC 6266), \citealt{Pooley02b} (NGC 6440), (3) \citealt{Heinke06b} (Terzan 5).}
\end{table*}

\section{Results}\label{sec_result}
\subsection{Hydrogen column density}\label{sec_nh}
We constrained the hydrogen column density ($N_H$) for each GC using our sample of LMXBs. In our primary analysis in this paper, we forced the N$_H$ to a single value for all observations of each source. We used the Tuebingen-Boulder ISM absorption model (TBABS) with \citet{Wilms00} abundances and \citet{BalucinskaChurch92} cross sections. (We tested the \citet{Verner96} cross sections, and found no difference in our results). We also tried using the abundances of \citet{Anders89} for comparison (see below). 
After constraining N$_H$ values for each target, we calculated the mean value for sources in each cluster. (We excluded Terzan 5 CX 9 from this analysis due to possible intrinsic absorption--see \S \ref{sec_terzan5}.) We compare these values with those produced by adopting reddening values, E(B--V), from the Harris catalog \citep[][2010 revision]{Harris96} and the relation between N$_H$ and optical extinction (A$_V$) from \citealt{Guver09}): 
\begin{equation}
N_H ({\rm cm}^{-2}) = (2.21\pm0.09)\times10^{21} A_V 
\end{equation}
(Table \ref{tab_nh}). The N$_H$ values calculated from $A_V$ using the relation of \citet{Guver09} agree nicely with fits to the X-ray spectra using the \citet{Anders89} abundances--which makes sense, as \citet{Guver09} used N$_H$ estimates from the literature on X-ray studies of supernova remnants, most of which used \citet{Anders89} abundances. However, N$_H$ values measured from spectral fitting using \citet{Wilms00} abundances are $\sim$27 \% higher than predicted using \citet{Guver09}.  This makes sense, because the \citet{Wilms00} abundances are typically $\sim$30\% lower for X-ray relevant elements. We therefore introduce a new relation between $N_H$ and $A_V$, designed specifically for use with the \citet{Wilms00} abundances, 
\begin{equation}
N_H ({\rm cm}^{-2}) = (2.81\pm0.13)\times10^{21} A_V 
\end{equation}
where the error is the 1-$\sigma$ scatter we measure in the relation from the fit. We derived this equation from fitting a linear model to $N_H$ values measured for 11 out of 12 sources in our sample (we excluded Terzan 5 CX 9 due to possible presence of enhanced absorption). We also investigated the effects of including dust scattering for point sources by adding a scattering model \citep{Predehl03} to our spectral model, with $A_V$ fixed to 0.117$\times N_H$/($10^{22}$ cm$^{-2}$). We found that dust scattering causes a slight decrease of $\approx$1\% in the $N_H$ values, and thus increases the $A_V$-to-$N_H$ conversion factor to 2.84$\times$10$^{21}$ cm$^{-2}$. As this work was being finalized, \citet{Foight15} published their investigation of the $A_V$-to-$N_H$ correlation using optical vs. X-ray studies of supernova remnants, also using \citet{Wilms00} abundances, which found the conversion factor to be 2.87$\pm0.12\times10^{21}$ cm$^{-2}$. This is in complete agreement with our results.

\begin{table*}
\centering
\begin{tabular}{@{}lllcccc@{}}
\hline
GC			&	E(B--V)	&	A$_V$	&Calculated N$_H$ (Guver09)	&Measured N$_H$ (Anders89)	&Measured N$_H$ (Wilms00)	&	Discrepancy	\\ 
			&	(mag)	&	(mag)	&($\times10^{21}$ cm$^{-2}$)	&($\times10^{21}$ cm$^{-2}$)	&($\times10^{21}$ cm$^{-2}$)	&	 (\%)			\\
\hline	
NGC 6266	&	0.47		&	1.46		&	3.2$\pm$0.2			&	3.2$\pm$0.4			&	4.1$\pm$0.4			&	22$\pm$9		\\
NGC 6440	&	1.07		&	3.32		&	7.3$\pm$0.5			&	7.2$\pm$0.6			&	10$\pm$1				&	27$\pm$9		\\
Terzan 5		&	2.61		&	8.09		&	18$\pm$1				&	17$\pm$2				&	 26$\pm$2			&	31$\pm$9		\\
\hline
\end{tabular}
\caption{Comparing $N_H$ for GCs based on different assumptions. Calculated $N_H$ is calculated based on relation between A$_V$ and N$_H$ provided by \citet{Guver09}. The uncertainties reported for calculated N$_H$ values based on \citet{Guver09} are only statistical uncertainties from the correlation. Measured N$_H$ values are based on X-ray spectroscopy of the sources in our sample (average and uncertainty calculated for each GC; see text) with two different assumptions for the abundances of elements (\citealt{Anders89} and \citealt{Wilms00}). E(B--V) values are obtained from the Harris Catalog \citep{Harris96}; to obtain $A_V$, we assumed R$_V$=3.1. Discrepancy is calculated between the measured N$_H$ using \citet{Wilms00} abundances and the calculated values based on the relation of \citet{Guver09}. All uncertainties are 90\% confidence.}
\label{tab_nh}
\end{table*}

\subsection{NGC 6266}\label{sec_ngc6266}
\begin{figure*}
\begin{center}
\includegraphics[scale=0.8]{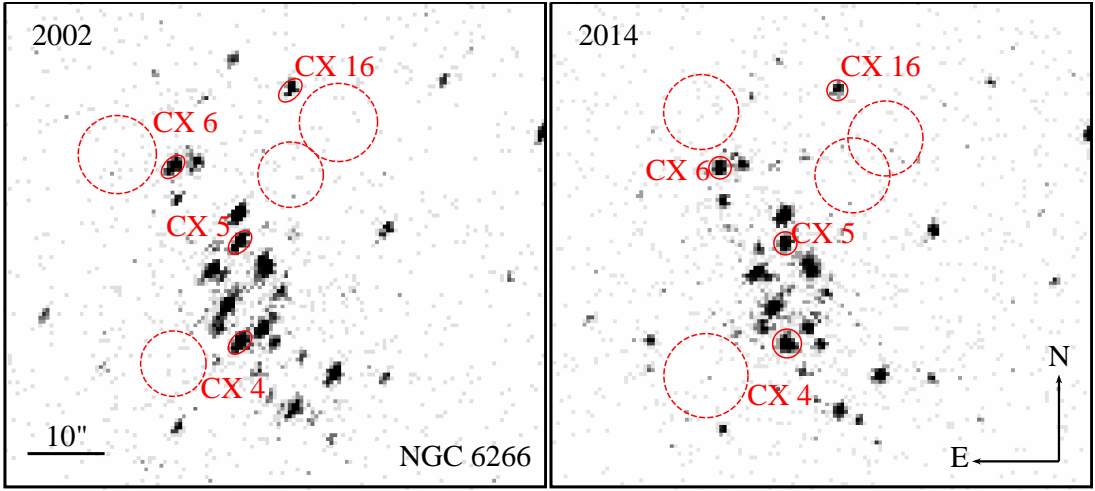}
\caption{X-ray (0.3-7 keV) images of NGC 6266 as seen by Chandra-ACIS in 2002 (left) and 2014 (right). Source extraction regions are represented by solid circles/ellipses, and background extraction regions are represented by dashed circles. In the 2002 observation, the centre of NGC 6266 was located 2.5 arcminutes off-axis. To account for the distortion of the point-spread function this offset induced, we used elliptical extraction regions for that observation.}
\label{fig_ngc6266}
\end{center}
\end{figure*}

NGC 6266 (M 62) is a massive dense globular cluster. At least 26 X-ray sources can be identified in {\it Chandra} images of the cluster, of which 5 have been suggested to be qLMXBs on the basis of their soft spectra and X-ray luminosities \citep{Pooley03,Heinke03d}.  One of these (the brightest, and with the hardest X-ray spectrum; reasonably fit by an absorbed powerlaw of photon index 2.5$\pm$0.1) shows strong evidence (from its radio/X-ray flux ratio) for containing a black hole \citep{Chomiuk13}.
We use two deep \chandraacis\ observations (Table \ref{tab_data}) to study the other 4 soft candidate qLMXBs, which we name CX 4, CX 5, CX 6, and CX 16 (ordering the detected sources by {\it Chandra} countrates). In the 2002 observation, the cluster core is 2.5 arcminutes off-axis, and the sources have visibly elongated point-spread functions. To address this issue, we used elliptical extraction regions of the same size for all sources. The 2014 observation was performed on-axis, so we used circular extraction regions. We show images of these observations and extraction regions in Fig. \ref{fig_ngc6266}. We ran \emph{wavdetect} to determine accurate positions and brightnesses of these X-ray sources, and corrected the astrometry by applying coordinate shifts to match the radio and X-ray positions of M62 CX1. 

We assumed a distance of 6.8 kpc, and a reddening of E(B--V)=0.47 mag, for NGC 6266 in our analysis \citep[][2010 revision]{Harris96}. Assuming R$_V$=3.1, this reddening gives A$_V$=1.46 mag. The $A_V$/$N_H$ relation of \citet{Guver09} which is based on \citet{Anders89} abundances, predicts N$_H$=3.2$\times$10$^{21}$ cm$^{-2}$ for this cluster based on \citet{Anders89} abundances; in \S \ref{sec_nh}, we use N$_H$ values measured from our spectral analysis of X-ray observations to derive a new relationship between $N_H$ and $A_V$, $N_H$=$2.8\times10^{21}$$A_V$, specifically for use with the \citet{Wilms00} abundances.

\begin{figure*}
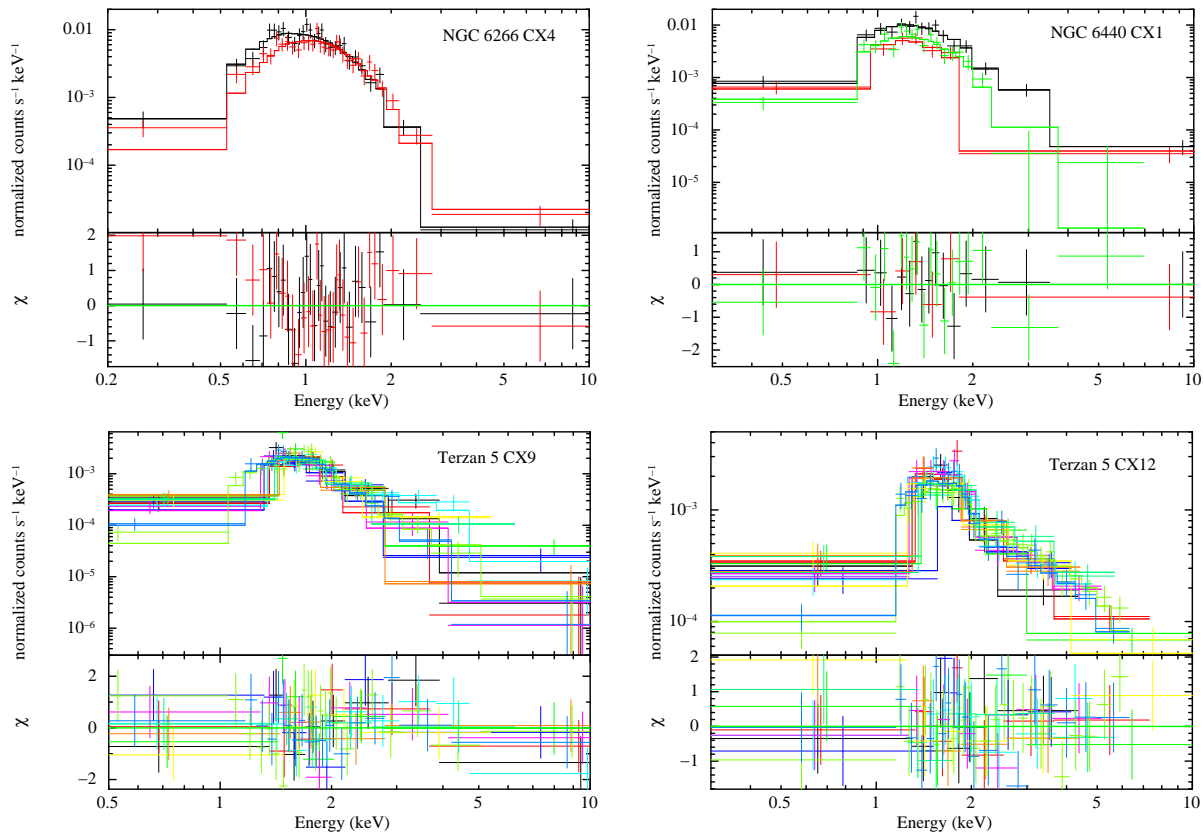

\begin{center}
\includegraphics[scale=0.3,angle=-90]{spec1_ngc6266_cx4b.eps}
\includegraphics[scale=0.3,angle=-90]{spec1_ngc6440_cx1b.eps}
\includegraphics[scale=0.3,angle=-90]{spec1_terzan5_cx9b.eps}
\includegraphics[scale=0.3,angle=-90]{spec1_terzan5_cx12b.eps}
\caption{Examples of extracted spectra; top panels show data (crosses) and model (line). Different colours represent different epochs. Note that in addition to changes in the source, changes in the detector may alter the observed spectrum (these changes are included in the model). Top left: NGC 6266 CX4 (see \S \ref{sec_ngc6266}). Only two \chandraacis\ epochs are available, but the spectral quality is high. Top right: NGC 6440 CX1 (see \S \ref{sec_ngc6440}). The black (2000) spectrum is brighter at all energies, particularly at higher energies; we attribute this to a difference in the powerlaw component, and possibly the thermal component.  Bottom left \& right: Terzan 5 CX9 \& Terzan 5 CX12 respectively, including spectra from 11 epochs (see \S \ref{sec_terzan5}). The spectral fit for CX9 shown here uses a hydrogen atmosphere model.  Note that there are still clear waves in the residuals (down at 2 keV, up at 3, down again at 5 keV), indicating problems with the fit (though the $\chi^2$ is formally acceptable, 87.42 for 87 dof). These residual patterns are not completely eliminated in a helium atmosphere model, though they are reduced. CX12 shows evidence for a strong powerlaw component in this fit, as well as evidence for variation in soft X-rays (particularly in the 4th observation, taken in April 2011, colored dark blue here).}
\label{fig_spec}
\end{center}
\end{figure*}

Due to the relatively low interstellar absorption and reasonable fluxes of our targets, the spectra from this cluster are of high quality (Fig. \ref{fig_spec}, top left). We found no evidence of temperature variations in these objects (Table \ref{tab_ngc6266}). For three of our targets (CX4, CX5 \& CX6) we constrain temperature variations to less than 5\%. For CX16 this limit is $<10$\% due to its faintness.

NGC 6266 CX4 is the only candidate qLMXB in NGC 6266 that shows evidence for a powerlaw component (Table \ref{tab_ngc6266}). To constrain the contribution from the powerlaw, we fixed the photon index to 1.5 due to the limited spectral quality at higher energies. We notice that there is a hint of a faint source $\sim 1''$ from CX4 (southwest of CX4, visible when the image is over-binned). To address this, we performed a second set of spectral analyses with a smaller source extraction region excluding the region of the possible nearby source, but we found no significant spectral changes suggesting the non-thermal emission is therefore intrinsic to CX4 (Table \ref{tab_ngc6266} contains fits performed with the standard extraction regions). 

We searched for short ($\sim$1000 s) timescale temporal variations in each observation by performing chi-squared and KS tests, and found no evidence of variation for any of our targets in NGC 6266. No source shows probability of constancy less than 0.06\% which is the 95\% confidence range when considering all trials (80, for all sources in this study).

\begin{table*}
\centering
\begin{tabular}{@{}clllll@{}}
\hline
Source			&N$_H$ ($10^{21}$ cm$^{-2}$)	&	Year		&	logT (K)				&	\% Variation	&	\% Powerlaw flux fraction (0.5-10 keV)	\\
\hline
NGC 6266 CX 4	&	3.9$\pm$0.4			&	2002		&	6.057$\pm$0.007		&	--			&	8$\pm$7					\\
Powerlaw $\Gamma$=($1.5$)	&				&	2014		&	6.056$\pm$0.006		&	$<3.2$		&	12$\pm$5					\\
$\chi_\nu^2$/d.o.f = 0.87/56	\\
\\	
NGC 6266 CX 5	&	3.9$\pm$0.4			&	2002		&	6.055$\pm$0.007		&	--			&	$<9$						\\
Powerlaw $\Gamma$=($1.5$)	&				&	2014		&	6.062$\pm$0.006		&	$<4.7$		&	$<7$						\\
$\chi_\nu^2$/d.o.f = 0.87/52	\\
\\
NGC 6266 CX 6	&	4.2$\pm$0.4			&	2002		&	6.056$\pm$0.006		&	--			&	$<9$						\\
Powerlaw $\Gamma$=($1.5$)	&				&	2014		&6.053$_{-0.007}^{+0.006}$	&	$<3.6$		&	$<7$						\\
$\chi_\nu^2$/d.o.f = 0.64/49	\\
\\
NGC 6266 CX 16	&	4.3$\pm$0.6			&	2002		&	5.96$\pm$0.01			&	--			&	$<23$					\\
Powerlaw $\Gamma$=($1.5$)&					&	2014		&	5.94$\pm$0.01			&	$<9$			&	$<15$					\\
$\chi_\nu^2$/d.o.f = 1.21/14	\\
\hline
\end{tabular}
\caption{NGC 6266: Results of spectral analysis using an absorbed NS atmosphere (TBABS*NSATMOS), plus a powerlaw added when needed. All uncertainties are 90$\%$ confidence. Temperature variations are calculated from the first observation in 2002. Powerlaw fractions are fractions of the total 0.5-10 keV flux.}
\label{tab_ngc6266}
\end{table*}

\subsection{NGC 6440}\label{sec_ngc6440}
NGC 6440 is a moderately extinguished GC near the Galactic center. There are at least 25 X-ray sources identified in this cluster, of which two are known transient LMXBs (NGC 6440 CX 1 = SAX J1748.9-2021, \citealt{intZand99,intZand01,Pooley02b, Bozzo15}; NGC 6440 X-2, \citealt{Altamirano10,Heinke10}). This cluster is 8.5 kpc away and has a reddening of E(B-V)=1.07 \citep{Harris96}. We used three long \chandraacis\ observations of this cluster, obtained in 2000, 2003, and 2009. We study the four brightest identified qLMXB candidates in or near the core of this cluster, CX 1, CX 2 (not to be confused with the AMXP denoted as NGC 6440 X-2, which is not detected in deep quiescence, e.g. \citealt{Heinke10}), CX 3, and CX 5, which were identified as likely qLMXBs by \citet{Pooley02b}.

\begin{figure*}
\begin{center}
\includegraphics[scale=0.7]{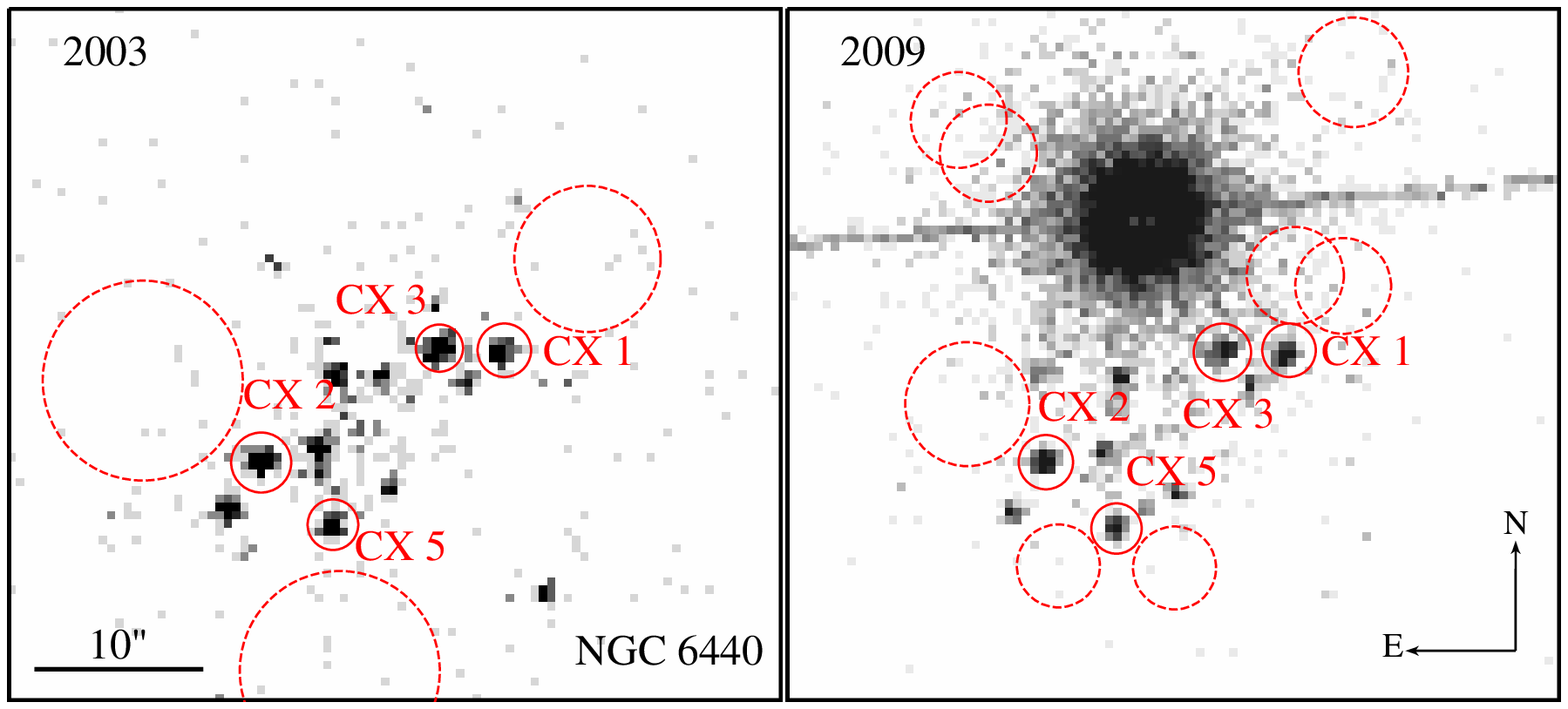}
\caption{NGC 6440 as seen by Chandra-ACIS in the 0.3-7 keV band in 2003 (left) and 2009 (right). Solid circles represent the source extraction regions and dashed circles show the selected background regions. In the 2009 observation, the image is contaminated by an outburst from NGC 6440 X-2 \citep{Heinke10}. We address this contamination for affected sources by choosing multiple background extraction regions at the same distance from X-2 as the target.}
\label{fig_ngc6440}
\end{center}
\end{figure*}

The \chandraacis\ observation of NGC 6440 in 2009 occurred while NGC 6440 X-2 was in outburst \citep{Heinke10} and the image is severely piled-up in the vicinity of this source. However, we were nevertheless able to use this observation for spectral analysis of qLMXBs, as NGC 6440 X-2 is located further away from the core of the cluster ($\sim 12''$) and the outburst was relatively faint (peaking at $\sim 1\times10^{36}$ erg s$^{-1}$). We addressed the contamination from the X-2 outburst on affected sources (CX 1, CX 2 \& CX 3) by extracting background spectra from multiple regions at the same angular distance as the target from X-2. Our targets and chosen extraction regions are shown in Fig. \ref{fig_ngc6440}. 

Two of the sources in this cluster (CX 1 and CX 3) show excesses at high energies to the NSATMOS fits, so we added a powerlaw component to their spectral models. In both cases, we froze the powerlaw photon index to the value found in the best fit to all three observations. We also note that there is a hint for the presence of a very faint source near CX 3 ($\sim$10 times fainter than CX 3, south of it, visible when the image is over-binned), however it does not affect the spectrum. The results of our spectral analyses for NGC 6440 are tabulated in Table \ref{tab_ngc6440}. \citet{Walsh15} measure powerlaw contributions consistent with our results for CX 2 and CX 3 in both 2003 and 2009 observations; We note that our extraction regions might differ from the ones chosen by \citet{Walsh15} as they do not discuss their choice of extraction regions. This is specially important due to the large and spatially varying background from the outburst of NGC 6440 X-2.

From our sample, CX 2 and CX 3 show no evidence of variation in the NS temperature. This is in general accord with the results of \citet{Walsh15}. These authors include a short (4 ks) \chandraacis\ observation of NGC 6440, 12 days after our 3rd observation, and find that the flux increases in this observation for CX 5.  If the spectrum is assumed to consist of only a NS atmosphere model, then the NS temperature must have increased in CX 5.  However, \citet{Walsh15} note that, since CX 5's spectrum may contain a powerlaw in this observation, the variability cannot be conclusively attributed to changes in the NS temperature. We find no evidence for variation in the spectrum of CX 5 between 2000 and 2009. We also detect the presence of a weak powerlaw component in the 2000 epoch at 1.7 $\sigma$. 

We found possible evidence for thermal variations in CX 1 in quiescence (Fig. \ref{fig_spec}, top right). We significantly detect a powerlaw with photon index 1.7$\pm2$ in the 2000 observation, but not in the 2003 or 2009 observations. Fixing this powerlaw index to the best-fit index, our best fit finds a $-10^{+7}_{-6}$\% change in temperature from 2000 to 2003, with the 2009 observation consistent with the 2003 observation (Table \ref{tab_ngc6440}). This marginal evidence for variation suggests the presence of continuing accretion during the 2000 observation (plausible considering that CX 1 erupted into outburst two years before, and one year after, the 2000 observation, \citealt{intZand01}).  

\citet{Cackett05} investigated the spectral variations of CX 1 using the first two \chandraacis\ observations of NGC 6440, but they argued that the observed spectral variations were likely due to changes in the powerlaw component exclusively.
 Cackett et al's best fit uses a powerlaw component with photon index 2.5($\pm1.0$), which explains the difference with our best fit (photon index 1.7$\pm2$); a softer powerlaw can explain flux differences in the soft X-rays, whereas our harder value of photon index require changes in the thermal component to explain. 
\citet{Walsh15} also fit the spectra of CX1 in these 3 observations (adding the short 2009 observation mentioned above) using C-stats in Xspec, find a different value from \citet{Cackett05} and basically agree with our results--the powerlaw photon index is measured at 1.6$\pm1.0$, and the 2000 data indicate a slightly higher NS temperature compared to the 2003 data, at 1$\sigma$ confidence (though not at 99\% confidence).  Walsh et al. also conclude (from $\chi^2$ fitting) that it is not clear whether the variation in CX1 was due exclusively to changes in the powerlaw component, exclusively to changes in the thermal component, or to changes in both. We agree with Walsh et al. that it is not clear whether the thermal component varied between the 2000 observation and later observations.

Searching for short ($\sim 1000 s$) timescale variations in each observation for all sources, we found no evidence of variation. No source shows probability of constancy less than 0.06\% which is the 95\% confidence range when considering all trials (80, for all sources in this study).

\begin{table*}
\centering

\begin{tabular}{@{}cllllll@{}}
\hline
Source			&N$_H$ ($10^{22}$ cm$^{-2}$)	&	Year		&	logT (K)				&	\% Variation	&	\% Powerlaw flux fraction (0.5-10 keV)	\\
\hline
NGC 6440 CX1		&	1.1$\pm$0.1			&	2000		&	6.16$\pm$0.01			&	10$_{-6}^{+7}$	&	26$\pm$10			\\
Powerlaw $\Gamma$=($1.7$)	&				&	2003		&	6.12$_{-0.02}^{+0.01}$	&	--			&	$<10$				\\
$\chi_\nu^2$/d.o.f = 0.84/37	&				&	2009		&	6.14$\pm$0.01			&	$<11$		&	$<8$					\\
\\
NGC 6440 CX2		&	1.1$\pm$0.1			&	2000		&	6.16$\pm$0.01			&	--			&	$<6$					\\
Powerlaw $\Gamma$=($1.5$)	&				&	2003		&	6.179$\pm$0.009		&	$<9$			&	$<9$					\\
$\chi_\nu^2$/d.o.f = 0.87/47	&				&	2009		&	6.172$\pm$0.007		&	$<7$			&	$<7$					\\
\\	
NGC 6440 CX3		&	0.9$\pm$0.1			&	2000		&	6.08$_{-0.03}^{+0.02}$	&	--			&	35$\pm$17			\\
Powerlaw $\Gamma$=($2$)	&				&	2003		&	6.10$\pm$0.02			&	$<17$		&	33$\pm$14			\\
$\chi_\nu^2$/d.o.f = 0.57/35	&				&	2009		&	6.10$\pm$0.02			&	$<17$		&	19$\pm$15			\\
\\
NGC 6440 CX5		&	1.0$\pm$0.1			&	2000		&	6.09$\pm$0.01			&	--			&	18$\pm$15			\\
Powerlaw $\Gamma$=($1.5$)	&				&	2003		&	6.10$_{-0.01}^{+0.02}$	&	$<10$		&	$<10$				\\
$\chi_\nu^2$/d.o.f = 1.00/25	&				&	2009		&	6.11$\pm$0.01			&	$<10$		&	$<7$					\\
\hline
\end{tabular}
\caption{Spectral analysis of qLMXBs in NGC 6440. We fit the spectra with an absorbed NS atmosphere (TBABS*NSATMOS), and included a powerlaw component when necessary. All uncertainties are 90$\%$ confidence. Temperature variations are calculated based on the coldest measured temperature for each source. Powerlaw fractions are fractions of the total 0.5-10 keV flux. In cases where the spectral quality allows us to constrain the powerlaw photon index (CX1 and CX3), we tied the index between observations and searched for the best-fit value, then froze the index to that best-fit value; otherwise we fixed the index to 1.5. The photon index used is listed in column 1.}
\label{tab_ngc6440}
\end{table*}

\subsection{Terzan 5}\label{sec_terzan5}
Terzan 5 is a massive, highly extincted GC near the Galactic center. It is located at a distance of 5.9$\pm0.5$ kpc \citep{Valenti07} and it harbours more than 40 X-ray sources \citep{Heinke06b}. There are three known transient LMXBs in this cluster, the most in any Galactic GC \citep{Bahramian14}. Terzan 5 has shown numerous outbursts (an overview up to 2012 is given in \citealt{Degenaar12}), several of which were not accurately localized (the most recent detected outburst without accurate localization being in 2002). It is therefore quite possible that other sources have been in outburst in the recent past.

 We use all available \chandraacis\ observations of this GC in which all sources are quiescent to study the brighter thermally-dominated qLMXBs without observed outbursts.  We start with the sample of candidate qLMXBs from \citet{Heinke06a}, which identified thermally-dominated qLMXBs by spectral fitting, or by hardness ratio plus inferred X-ray luminosity.  Analyses of the quiescent behaviour of the outbursting sources Ter 5 X-1 (EXO 1745-248), X-2 (IGR J17480-2446), and X-3 (Swift J174805.3-244637) have been reported in \citealt{Degenaar11a,Degenaar11b, Degenaar13, Bahramian14, Degenaar15}, so we exclude those sources here. 

For the purpose of imaging and source detection, we merge all available observations following CIAO threads\footnote{http://cxc.harvard.edu/ciao/threads/combine/}. We corrected relative astrometry in all frames by matching the coordinates of 6 bright sources. We ran \emph{reproject\_aspect} to create a new reprojected aspect solution for each observation. Then we applied these aspect solutions to the event files by running \emph{reproject\_events}, and finally we combined the reprojected event files by running \emph{reproject\_obs}. 

\begin{figure*}
\begin{center}
\includegraphics[scale=0.6]{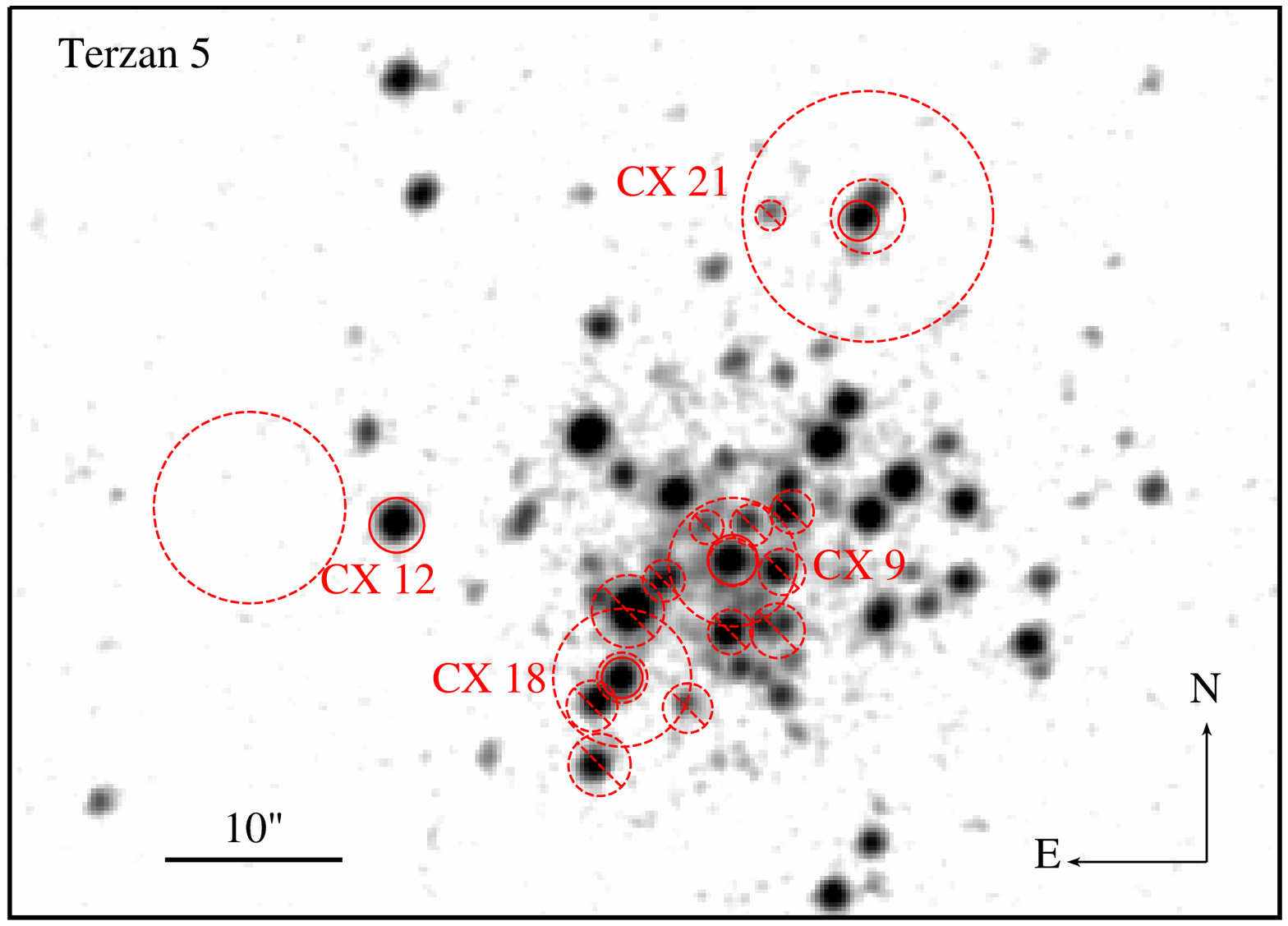}
\caption{Stacked and smoothed \chandraacis\ image of Terzan 5 in 0.3-7 keV band (total $\sim$620 ks). Source extraction regions for our targets are shown with red solid circles, and regions for background regions are represented by dashed circles/annuli. For targets in complex regions, we extracted background from annuli. Dashed circles with slashes represent excluded regions in background extraction (i.e. contaminating sources).}
\label{fig_terzan5}
\end{center}
\end{figure*}

We selected the candidate qLMXBs which had sufficient counts for spectral analysis in each epoch ($\geq60$; i.e. more than 4 spectral bins of 15 counts), leaving us with a sample of four thermally-dominated qLMXBs, identified as CX9, CX12, CX18, and CX21 in \citet{Heinke06a}.  We used the merged image to choose source and background spectral extraction regions. This is important since in cases like CX9 and CX18, the vicinity of the source is complex and there can be multiple sources of contamination. In cases where the vicinity of the source is complex/crowded, we extracted background spectra from annuli around the targets, excluding detected sources. The combined image of Terzan 5, with our extraction regions, is presented in Fig. \ref{fig_terzan5}. 

We combined spectra from observations which occurred less than a month apart (using \emph{combine\_spectra}), to obtain better statistics. Investigating these observations individually, no spectral variation was detected. These observations are marked in Table \ref{tab_data}. All our targets in Terzan 5 show a high energy excess suggesting the presence of a powerlaw component in their spectra with average fraction of $\sim$10$\%$ of the total flux for CX9, CX18, and CX21 and an average fraction of $\sim$33$\%$ for CX 12. Comparing these with our targets in NGC 6266 and NGC 6440, where the powerlaw component is mostly detected when flux fraction is $\geq$10$\%$, suggests that our detection of a powerlaw component in all our targets in Terzan 5 might be due to the greater depth of these observations, combined with our relatively higher sensitivity to hard vs. soft photons in this heavily absorbed cluster. We froze the powerlaw photon index to 1.5 for CX9, CX18 and CX21, due to a lack of sufficient signal to constrain it. For CX12 we freeze it to the value of 1.8, obtained from the best fit to all observations.

The values of N$_H$ for CX12, CX18 and CX21 are all in agreement with the value of 2.6$\pm0.1\times 10^{22}$ cm$^{-2}$ from previous studies (see \S \ref{sec_nh})\footnote{In many previous studies, abundances from \citet{Anders89} have been used for absorption models. Using these abundances, \citet{Bahramian14} find N$_H=1.74_{-0.08}^{+0.06}\times10^{22}$ cm$^{-2}$. However in the same paper they show that using \citet{Wilms00} abundances gives an equivalent value of 2.6$\pm0.1\times 10^{22}$ cm$^{-2}$.}. However, CX9 has a hydrogen column density of 3.1$\pm0.2\times 10^{22}$ cm$^{-2}$, higher than the measured value for the cluster. We investigate the high-resolution reddening map of Terzan 5 (\citealt{Massari12}, available online from Cosmic-Lab \footnote{http://www.cosmic-lab.eu/tred/tred.php}), and find no significant difference ($<$1.5\%) in reddening between the direction of CX9 and the rest of the sources in our sample. This suggests that the inferred high N$_H$ value may be caused by intrinsic absorption in the system. Spectra of CX9 are shown in Fig. \ref{fig_spec} (bottom left).

We also performed spectral fits on CX9 by replacing the non-magnetic hydrogen atmosphere model (NSATMOS) with a non-magnetic helium atmosphere model (NSX in Xspec, \citealt{Ho09}). This fit gives a slightly better  $\chi^2$ compared to the hydrogen atmosphere model ($\chi^2$ decreased from 114 to 109, where both cases have 102 degrees of freedom), and also a lower absorption value of $N_H=2.9\pm0.2\times10^{22}$ cm$^{-2}$, consistent with the rest of the cluster (Table \ref{tab_ter5cx9}). These results indicate that a helium atmosphere is a possible explanation for the unusual aspects of CX9's spectrum (see \S \ref{sec_disc}). Note that there are still clear waves in the residuals (down at 2 keV, up at 3, down again at 5 keV), indicating problems with the fit (though the $\chi^2$ is formally acceptable, 87.42 for 87 dof). These residual patterns are not completely eliminated in a helium atmosphere model, though they are reduced.

\begin{table}
\centering
\begin{tabular}{@{}lcccccccc@{}}
\hline
\multicolumn{4}{c}{Terzan 5 CX9}\\
\hline
				&\multicolumn{2}{c}{Hydrogen}				&	Helium		\\
Parameter			&	Trial 1	 			&	Trial 2	&	Trial 3			\\
\hline
N$_H$ ($10^{22}$cm$^{-2}$)&	3.1$\pm0.2$		&	(2.6)		&	2.9$\pm0.2$	\\
logT	(K)			&	6.12$\pm$0.01			&6.21$\pm$0.06&	6.10$\pm$0.01	\\
NS R (km)			&	(10)					&	7$\pm$1	&	(10)			\\
PL flux $^A$	&1.1$\pm${0.3}					&	1.3$\pm$0.3&	0.9$\pm$0.3	\\
$\chi_\nu^2$/d.o.f	&	1.11/102				&	1.24/102	&	1.07/102		\\
NHP  	&	0.20					&	0.05		&	0.29			\\
\hline
\end{tabular} 
\caption{Tests of hydrogen (NSATMOS) and helium (NSX) atmosphere models for Terzan 5 CX9. Two trials use hydrogen atmospheres: in trial 1, N$_H$ is free, while the NS radius is fixed to 10 km. In trial 2, N$_H$ is fixed to the average value for Terzan 5, and the NS radius is free. Trial 3 is the same as trial 1, except using a helium atmosphere. Since we didn't find any sign of variation in our initial fits, all values are tied between epochs. The neutron star mass is assumed to be 1.4 M$\odot$ in each case. Uncertainties are 90\% confidence, and values in parentheses are frozen.  $A$: powerlaw fluxes are in units of 10$^{-14}$erg s$^{-1}$ cm$^{-2}$, in the 0.5-10 keV range. NHP is null hypothesis probability.}
\label{tab_ter5cx9}
\end{table}

We found no evidence regarding spectral variations between observations in CX 9, CX 18 and CX 21. However CX 12 shows marginal evidence of variation over time (Table \ref{tab_terzan5}). Our main fit, in which only the NS temperature and powerlaw flux are free among observations, suggests that these variations are caused by changes in the temperature of the NS, particularly a decrease in the 2011-4 epoch, and to a lesser degree in the two 2013 observations.  However, the substantial powerlaw component, which is particularly strong in the epochs with the coldest NS temperature measurements, suggests that subtle changes in the powerlaw (both flux and index) might cause the variations (Fig. \ref{fig_spec}, bottom right). We note that thermal variations are only suggested in 2 epochs (2011-4 and 2013-2) out of 11. However we get an acceptable chi-squared when fitting the temperatures to a straight line. Thus we agree with \citet{Walsh15} that these variations are not statistically significant. 

To investigate the origin of CX12's spectral variations more carefully, we perform fits letting different pairs of model parameters vary; kT and PL flux (Fit 1), N$_H$ and PL flux (Fit 2), or PL $\Gamma$ and PL flux (Fit 3; see Table \ref{tab_ter5cx12}).  In fit 1, variations in both kT and the PL flux are seen, and the fit is reasonable.  However, the variations in kT appear to be anticorrelated with the variations in PL flux, which is unlike, for instance, the behaviour of Cen X-4 \citep{Cackett10}. It should be noted that the signature of variation comes mainly from two data points (epochs 2011/4 and 2013/2). Performing a chi-squared test of constancy on temperature measured in all epochs gives a null hypothesis probability of 0.55, providing little evidence for variation.
Compared to fit 1, fit 3 shows an increase of 14 in $\chi^2$ (107 compared to 93, both fits have 96 degrees of freedom, giving a relative likelihood of 9$\times10^{-4}$ for fit 3 compared to fit 1 based on Akaike information criterion), indicating that changes in the powerlaw alone (both index and flux) are insufficient to drive the variations. However, fit 2, varying N$_H$ and PL flux, gave a similar fit quality to fit 1 ($\chi^2=92.6$ compared to $\chi^2=93.4$, both with 96 degrees of freedom). 
Thus, the variations may plausibly be caused by changes in the NS temperature, or changes in intrinsic absorption, along with changes in the PL flux. 
Since we do not have independent evidence of the inclination of CX12, it is quite possible that this system is nearly edge-on, and that we suffer varying obscuration levels at different times.  

\begin{figure*}
\begin{center}
\includegraphics[scale=0.4]{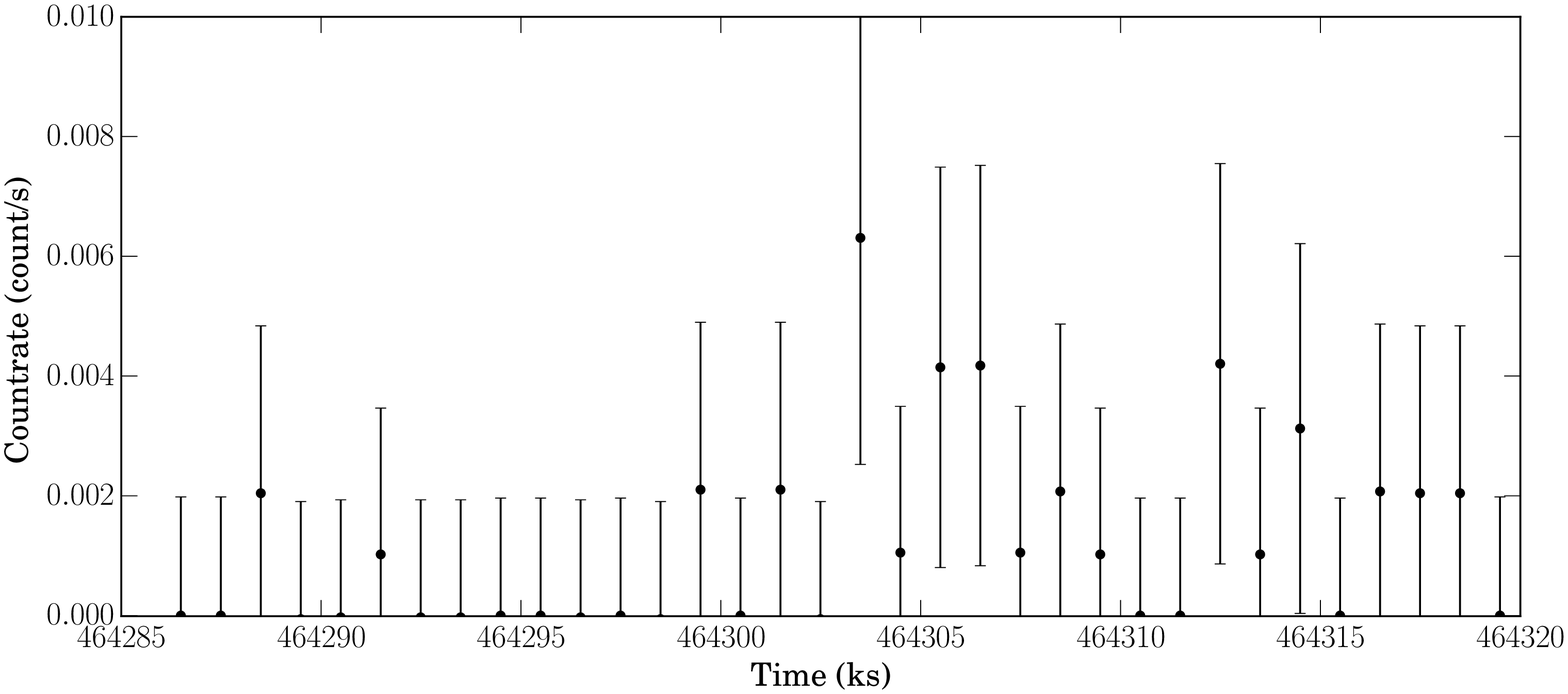}
\caption{Binned lightcurve of Terzan 5 CX 21 from observation 14475 (Sept. 2012). A KS test gives probability of constancy = 4.6$\times10^{-4}$ for this lightcurve, which is slightly higher than 95\% confidence, when accounting for the number of trials. It is possible that this variation may be due to eclipse. However we do not find strong evidence for eclipse in other epochs of this source.}
\label{fig_terzan5cx21lc}
\end{center}
\end{figure*}

We performed chi-squared and KS tests on individual lightcurves extracted from each observation for our target sources (excluding observation 13705, due to its short exposure). We found evidence of short ($\sim$1000 s) timescale variations in Terzan 5 CX 21's lightcurve, in observation 14475 (Sept. 2012, Fig. \ref{fig_terzan5cx21lc}). A KS test gives probability of constancy = 4.6$\times10^{-4}$ for this lightcurve, which is slightly higher than 95\% confidence, when accounting for the number of trials (80, for all sources in this study). Note that although there is evidence for presence of the PL in this epoch, the PL flux fraction is rather small (13$_{-11}^{+16}$ \%), suggesting that the thermal fraction is also likely to vary. It is also possible that variations observed for this source are due to eclipse in the system, however we do not find evidence for eclipse in other epochs. No other source shows KS probability of constancy less than $0.06 \%$ (95\% confidence).

\begin{table*}
\centering
\begin{tabular}{@{}cllll@{}}
\hline
Source		&	Epoch	&	logT (K)				&	\% Variation	&	\% Powerlaw flux fraction (0.5-10 keV)	\\
\hline
Terzan 5 CX 9	&	2003		&	6.13$_{-0.02}^{+0.01}$	&	$<$15		&	7$_{-6}^{+7}$		\\
Powerlaw $\Gamma$=(1.5)		&	2009		&	6.11$\pm$0.02			&	$<$12		&	$<$14			\\
$N_H = 3.1\pm0.2\times10^{22}$	&	2011-2	&	6.12$\pm$0.02			&	$<$15		&	12$_{-8}^{+10}$  	\\
$\chi_\nu^2$/d.o.f = 1.00/87		&	2011-4	&	6.11$\pm$0.02			&	$<$12		&	8$_{-7}^{+9}$ 		\\
			&	2011-9	&	6.10$\pm$0.02			&	--			&	22$_{-9}^{+11}$	\\
			&	2012-5	&	6.12$_{-0.02}^{+0.01}$	&	$<$12		&	$<$8				\\
			&	2012-9	&	6.11$\pm$0.02			&	$<$12		&	11$_{-8}^{+10}$	\\
			&	2012-10	&	6.13$\pm$0.02			&	$<$17		&	$<$8				\\
			&	2013-2	&6.125$_{-0.008}^{+0.007}$	&	$<$13		&	4$_{-2}^{+3}$		\\
			&	2013-7	&	6.12$\pm$0.02			&	$<$15		&	10$_{-8}^{+10}$	\\
			&	2014		&6.128$_{-0.009}^{+0.008}$	&	$<$14		&	$<$5				\\
\\	
Terzan 5 CX 12								&	2003		&	6.09$\pm$0.02			&23$_{-16}^{+31}$	&	14$_{-8}^{+10}$		\\
Powerlaw $\Gamma$=$1.8\pm$0.3	&	2009		&	6.08$_{-0.03}^{+0.02}$	&20$_{-18}^{+31}$	&	27$_{-10}^{+13}$		\\
$N_H = 2.6\pm0.2\times10^{22}$				&	2011-2	&	6.07$_{-0.03}^{+0.02}$	&	$<$48		&	29$_{-13}^{+16}$		\\
$\chi_\nu^2$/d.o.f = 0.97/96					&	2011-4	&	6.00$_{-0.08}^{+0.04}$	&	--			&	52$_{-18}^{+27}$		\\
			&	2011-9	&	6.09$_{-0.03}^{+0.02}$	&23$_{-18}^{+32}$	&	28$_{-9}^{+11}$		\\
			&	2012-5	&	6.09$_{-0.02}^{+0.02}$	&23$_{-16}^{+32}$	&	27$_{-9}^{+11}$		\\
			&	2012-9	&	6.07$_{-0.04}^{+0.03}$	&	$<$51		&	34$_{-13}^{+17}$		\\
			&	2012-10	&	6.08$_{-0.03}^{+0.02}$	&20$_{-18}^{+31}$	&	29$_{-13}^{+16}$		\\
			&	2013-2	&	6.04$_{-0.02}^{+0.02}$	&	$<$38		&	44$_{-7}^{+8}$			\\
			&	2013-7	&	6.05$_{-0.06}^{+0.03}$	&	$<$44		&	49$_{-16}^{+22}$		\\
			&	2014		&	6.07$_{-0.01}^{+0.02}$	&17$_{-12}^{+31}$	&	29$_{-6}^{+7}$			\\
\\		
Terzan 5 CX 18				&	2003		&	6.09$\pm$0.02	&	$<$17	&	13$_{-8}^{+11}$		\\
Powerlaw $\Gamma$=(1.5) 	&	2009		&	6.09$\pm$0.02	&	$<$17	&	$<$13				\\
$N_H = 2.7\pm0.2\times10^{22}$&	2011-2	&	6.10$\pm$0.02	&	$<$20	&	$<$7					\\
$\chi_\nu^2$/d.o.f = 1.12/51	&	2011-4	&	6.08$\pm$0.02	&	$<$15	&	$<$14				\\
			&	2011-9	&	6.08$\pm$0.02			&	$<$15		&	7$_{-6}^{+8}$			\\
			&	2012-5	&	6.07$\pm$0.02			&	$<$12		&	7$_{-7}^{+9}$			\\
			&	2012-9	&	6.07$_{-0.03}^{+0.02}$	&	--			&	$<$13				\\
			&	2012-10	&	6.09$_{-0.03}^{+0.02}$	&	$<$17		&	$<$13				\\
			&	2013-2	&	6.08$\pm$0.01			&	$<$12		&	12$\pm$1				\\
			&	2013-7	&	6.07$_{-0.03}^{+0.02}$	&	$<$12		&	10$_{-10}^{+14}$		\\
			&	2014		&	6.09$\pm$0.01			&	$<$15		&	7$\pm$1				\\
\\	
Terzan 5 CX 21	&	2003		&	6.04$_{-0.03}^{+0.02}$	&	$<$15		&	16$_{-12}^{+16}$		\\
Powerlaw $\Gamma$=(1.5) 		&	2009		&	6.06$_{-0.03}^{+0.02}$	&	$<$20		&	10$_{-9}^{+12}$		\\
$N_H = 2.4\pm0.2\times10^{22}$	&	2011-2	&	6.07$_{-0.03}^{+0.02}$	&	$<$23		&	$<$12				\\
$\chi_\nu^2$/d.o.f = 0.92/36		&	2011-4	&	6.03$\pm$0.03			&	--			&	17$_{-11}^{+16}$		\\
			&	2011-9	&	6.08$\pm$0.02			&	$<$26		&	$<$10				\\
			&	2012-5	&	6.04$\pm$0.02			&	$<$15		&	$<$15				\\
			&	2012-9	&	6.04$_{-0.04}^{+0.02}$	&	$<$15		&	13$_{-11}^{+16}$		\\
			&	2012-10	&	6.05$_{-0.03}^{+0.02}$	&	$<$17		&	11$_{-10}^{+14}$		\\
			&	2013-2	&	6.04$\pm$0.01			&	$<$12		&	11$_{-4}^{+5}$			\\
			&	2013-7	&	6.06$\pm$0.02			&	$<$20		&	$<$7					\\
			&	2014		&	6.05$\pm$0.01			&	$<$15		&	$<$16				\\
\\	
\hline
\end{tabular}
\caption{Spectral analyses of Terzan 5 targets with an absorbed NS atmosphere + powerlaw. All uncertainties are 90$\%$ confidence. Temperature variations are calculated based on the coldest measured temperature for each source.  The ``Epoch'' values give the year and month, in cases where multiple observations per year were performed. Powerlaw fractions are fractions of the total unabsorbed 0.5-10 keV flux.  All targets in Terzan 5 show significant evidence for the presence of a powerlaw component in their spectrum. In cases where the spectral quality allows us to constrain the powerlaw photon index (CX 12), we tied the index between observations and searched for the best-fit value, then froze the index to that best-fit value; otherwise we fixed the index to 1.5. The photon index used is listed in column 1. N$_H$ values are in units of cm$^{-2}$. }
\label{tab_terzan5}
\end{table*}

\begin{table*}
\centering
\begin{tabular}{@{}lcccccccc@{}}
\hline
\multicolumn{9}{c}{Terzan 5 CX12}\\
\hline
		&	\multicolumn{2}{c}{Fit 1}				&& 	\multicolumn{2}{c}{Fit 2}						&&	\multicolumn{2}{c}{Fit 3}				\\
		&\multicolumn{2}{c}{Constant N$_H$ \& $\Gamma$}	&& 	\multicolumn{2}{c}{Constant T$_{NS}$ \& $\Gamma$} &&	\multicolumn{2}{c}{Constant N$_H$ \& T$_{NS}$}				\\		
\cline{2-3}
\cline{5-6}
\cline{8-9}
		&	logT					&	PL flux					&&	N$_H$			&	PL flux					&&	$\Gamma$		&	PL flux 			\\
Epoche	&	(K)					&(10$^{-14}$erg s$^{-1}$ cm$^{-2}$)	&&(10$^{22}$cm$^{-2}$)	&(10$^{-14}$erg s$^{-1}$ cm$^{-2}$)	&&		&(10$^{-14}$erg s$^{-1}$ cm$^{-2}$)		\\
\hline
2003		&	6.09$\pm$0.02			&	2$\pm$1					&&	2.5$_{-0.2}^{+0.3}$	&	2$\pm$1					&&	3.1$_{-0.7}^{+0.8}$  	&	7$_{-3}^{+6}$		\\
2009		&	6.08$_{-0.03}^{+0.02}$	&	5$\pm$2					&&	2.5$_{-0.2}^{+0.3}$	&	5$\pm$1					&&	2.4$\pm$0.7    		&	7$_{-2}^{+4}$		\\
2011-2	&	6.07$_{-0.03}^{+0.02}$	&	4$\pm$2					&&	2.7$_{-0.3}^{+0.5}$	&	4$\pm$2					&&	2.3$\pm$0.7    		&	6$_{-2}^{+3}$		\\
2011-4	&	6.00$_{-0.08}^{+0.04}$	&	6$\pm$2					&&	3.5$_{-0.4}^{+0.6}$	&	5$\pm$2					&&	1.3$_{-0.7}^{+0.6}$  	&	4$\pm$1			\\
2011-9	&	6.09$_{-0.03}^{+0.02}$	&	5$\pm$2					&&	2.5$_{-0.2}^{+0.3}$	&	5$\pm$1					&&	2.4$\pm$0.6	     	&	8$_{-2}^{+3}$		\\
2012-5	&	6.09$_{-0.02}^{+0.02}$	&	5$\pm$1					&&	2.4$_{-0.2}^{+0.3}$	&	5$\pm$1					&&	2.4$\pm$0.6   		&	8$_{-2}^{+4}$		\\
2012-9	&	6.07$_{-0.04}^{+0.03}$	&	6$\pm$2					&&	2.5$_{-0.3}^{+0.5}$	&	5$\pm$2					&&	1.8$\pm$0.8   		&	6$_{-2}^{+3}$		\\
2012-10	&	6.08$_{-0.03}^{+0.02}$	&	5$\pm$2					&&	2.3$_{-0.3}^{+0.4}$	&	5$\pm$2					&&	2$\pm$1	  		&	7$_{-2}^{+5}$		\\
2013-2	&	6.04$_{-0.02}^{+0.02}$	&	7$\pm$2					&&	3.0$_{-0.2}^{+0.3}$	&	6$\pm$1					&&	1.6$\pm$0.3  	  	&	6$\pm$1			\\
2013-7	&	6.05$_{-0.06}^{+0.03}$	&	8$\pm$2					&&	2.8$_{-0.4}^{+0.6}$	&	8$\pm$2					&&	1.8$\pm$0.5   	  	&	8$\pm$2			\\
2014		&	6.07$_{-0.01}^{+0.02}$	&	5$\pm$1					&&	2.6$_{-0.2}^{+0.2}$	&	5$\pm$1					&&	2.2$\pm$0.4   	  	&	6$\pm$1			\\
\cline{2-3}
\cline{5-6}
\cline{8-9}
\\
$\chi_\nu^2$/d.o.f&	\multicolumn{2}{c}{0.97/96}						&&	\multicolumn{2}{c}{0.96/96}						&&		\multicolumn{2}{c}{1.12/96}			\\
NHP &	\multicolumn{2}{c}{0.55}						&&	\multicolumn{2}{c}{0.58}							&&		\multicolumn{2}{c}{0.20}				\\
\hline
\end{tabular}
\caption{Testing three possibilities to explain the spectral variations in Terzan 5 CX12. Fit 1: Fixing N$_H$ and the powerlaw photon index to their best-fit value across all spectra (N$_H=2.6\times10^{22}$cm$^{-2}$, $\Gamma=1.8$), while letting the NS temperature and powerlaw flux vary between epochs. Fit 2: Allowing N$_H$ and powerlaw flux to vary across epochs, while fixing the NS temperature and powerlaw photon-index ($\log T = 6.08 $, $\Gamma=1.7$). Fit 3: Only photon index and powerlaw flux are allowed to vary between epochs, with N$_H$ and the NS temperature frozen (N$_H=2.5\times10^{22}$cm$^{-2}$, $\log T = 6.05$). powerlaw fluxes are given in 0.5-10 keV band. The NS mass and radius are frozen to their canonical values of 1.4 M$_\odot$ and 10 km in all cases. All uncertainties are 90\% confidence. NHP is null hypothesis probability.}
\label{tab_ter5cx12}
\end{table*}

\section{Discussion}\label{sec_disc}
\subsection{Terzan 5 CX9: a NS with a helium atmosphere?}
Terzan 5 CX 9 shows an unusual spectral shape compared to other qLMXBs, with significantly lower flux at low energies (between 0.1 and 1 keV), compared to similar sources. This could be due to a higher amount of absorption. Investigating the reddening map of this cluster, we can rule out differential reddening. However, there could be additional intrinsic absorption, for instance if we are looking at the system edge-on.

Another possibility is that this NS may have an atmosphere composed of helium. Ultracompact X-ray binaries (with a white dwarf donor star, possessing no hydrogen) are fairly common among bright LMXBs in globular clusters \citep[e.g.][]{Zurek09}.  Helium white dwarf donors might thus be expected to lead to helium atmospheres on the NSs; there is evidence for low hydrogen content in X-ray bursts from some bright ultracompact X-ray binaries \citep{Cumming03,Galloway08}. The X-ray spectra of helium atmospheres are similar to hydrogen atmospheres, but slightly harder \citep{Romani87,Ho09}. An observed spectrum will thus give different parameter values for He vs. H atmospheres, and several papers have considered the importance of atmosphere composition for attempts to constrain the NS mass and radius \citep{Servillat12,Catuneanu13,Lattimer14,Heinke14}.  For relatively low-count spectra such as CX9, which parameter is different from other NS qLMXBs--e.g., atmospheric composition (H or He), vs. $N_H$--may not be possible to differentiate.  No helium atmosphere has yet been confidently identified on a NS, so any evidence in favor of a helium atmosphere is intriguing.

\subsection{Possible continuous accretion for NGC 6440 CX1 \& Terzan 5 CX12}
NGC 6440 CX 1 is the brightest X-ray source in the cluster. It has shown multiple X-ray outbursts over the last 20 years (1998, 2001, 2005, 2010, 2015; \citealt{intZand99,intZand01,Markwardt05,Patruno10, Homan15}). In quiescence, it shows a mostly thermal spectrum although there are signs of variations in the spectrum over the course of years. In the 2000 observation, we find that the spectrum requires a powerlaw component for a good fit, but in later observations this component is not required. \citet{Cackett05} argued that the changing normalization of the powerlaw component, with photon index$\sim$2.5, was the main source of spectral variation. However, we suggest that the unusually high photon index used there led to the overestimation of the  powerlaw's contribution. \citet{Walsh15} also fit the spectra of CX 1 with similar models. They find that formally, either the powerlaw or thermal component can be the source of variation, and we agree with that finding. However, when we hold the NS temperature constant across the epochs while allowing the powerlaw normalization to vary (with photon index tied to a single value across epochs), the best-fit value of the powerlaw photon index will increase to 2.7$\pm$0.5 (compared to 1.7$\pm$2 when the NS temperature also varies). This agrees with Walsh et al.'s fit results when assuming NS temperature fixed (the first fit in their Table 4), where they find a photon index of 3.1$\pm$0.5. The anomalously high photon index required for such a fit strongly argues that the thermal component may also be varying.

Terzan 5 CX12 is well-fit by an absorbed two-component (thermal+powerlaw) model, with an N$_H$ column typical of sources in the cluster (2.6$\pm$0.1$\times$10$^{22}$ cm$^{-2}$). CX12 shows hints of spectral variations, concentrated at low energies ($<2$ keV; see Figure 2, lower right). Thus it's unlikely that the powerlaw component is responsible for these variations. Intrinsic absorption (e.g. an edge-on system, where an uneven accretion disk occasionally blocks the line of sight), or continuing thermal variation due to accretion are possible origins of these variations. 

\subsection{Variability of thermal component in qLMXBs}
In our sample of 12 globular cluster qLMXBs, NGC 6440 CX1 and Terzan 5 CX12 are the only sources that show evidence of variations in the thermal component over timescales of years. In both cases, the thermal component appears to be the most plausible origin of the variations, but in neither case can thermal variations be proven.  For NGC 6440 CX1, the best-fit thermal variation is only a 10\% temperature change, while for Terzan 5 CX12 one observation appears to have a 20\% drop. 

Terzan 5 CX 21 shows evidence of variability on timescales of hours (within observations), although we did not detect variations on longer timescales (between observations). This variability may be due to variations in the (dominant) thermal component in quiescence.  Alternatively, it may be caused by eclipses by the companion, if the system is at a high inclination angle.

A crucial result of our analysis is that we have assembled an additional 7 qLMXBs (in addition to 4 others previously reported, see Introduction) with multiple high-quality X-ray spectra showing little or no evidence for a powerlaw component; NGC 6266 CX4, CX5, CX6, and CX16, and NGC 6440 CX1 (considering only the 2003 and 2009 observations), CX2, and CX5.  None of these objects show evidence for a powerlaw component comprising more than 12\% of the 0.5-10 keV flux (except for the 2000 observation of CX1, which we exclude here). The upper limits on a powerlaw (assuming a photon index of 1.5) are, in four cases, $<$10\% of the 0.5-10 keV flux, and $<$17\% or $<$23\% in two others.
Using {\it Chandra} observations spaced over 9-12 years (6 for NGC 6440 CX1), we can constrain the thermal emission from these NS to not have varied by more than 10\% in any of these sources.  Combined with the literature constraints on 4 other sources discussed in \S 1 (and in agreement with the recent work of \citealt{Walsh15}), this provides increasing evidence that the thermal X-ray spectral components of NS qLMXBs without powerlaw components to their spectra are not powered by continuing accretion. (Of course, a clear detection of intrinsic variation of the thermal component in a source without a powerlaw component would disprove this hypothesis.)  This contrasts with recent works that indicate that in some NS qLMXBs with powerlaw components, the powerlaw, as well as some of the thermal component, is powered by continuing, variable, accretion onto the NS surface  \citep{Cackett10,Bahramian14,Chakrabarty14,D'Angelo14,Wijnands14}.

This evidence against continuing accretion for those qLMXBs without powerlaw components also supports the use of these qLMXB spectra to obtain constraints on the mass and radius (and thus on the equation of state) of NSs, since a lack of accretion indicates that the atmosphere will not be contaminated by heavy elements.

\section*{Acknowledgments}
This research has made use of the following data and software packages: observations made by the Chandra X-ray Observatory, data obtained from the Chandra Data Archive, software provided by the Chandra X-ray Center (CXC) in the application package CIAO, and software provided by the High Energy Astrophysics Science Archive Research Center in the HEASOFT package. We acknowledge extensive use of the ADS and arXiv. AB thanks G.R.Sivakoff and T.J.Maccarone for helpful discussions. COH acknowledges financial support from NSERC Discovery Grants, an Alberta Ingenuity New Faculty Award, and an Alexander von Humboldt Fellowship. ND acknowledges support via an EU Marie Curie Intra-European fellowship under contract no. FP-PEOPLE-2013-IEF-627148. LC acknowledges support from a {\it Chandra} grant associated with Obs.ID \#15761. WCGH acknowledges support from STFC in the UK. This work was partially supported by the NSF through grant AST-1308124.

\bibliographystyle{mn2e}
\bibliography{ref_list}

\bsp

\label{lastpage}

\end{document}